\documentclass[aps,pre,groupedaddress,reprint]{revtex4-1}

\usepackage{amsmath,amssymb,amsfonts,graphicx}

\begin{document}

\title{Indirect identification of damage functions from damage records}

\author{J. Micha Steinh\"auser}
\affiliation{Potsdam Institute for Climate Impact Research -- 
14412 Potsdam, Germany, EU}
\affiliation{University of Oldenburg - 26129 Oldenburg, Germany, EU.}

\author{Diego Rybski}
\email[]{ca-dr@rybski.de}
\affiliation{Potsdam Institute for Climate Impact Research -- 
14412 Potsdam, Germany, EU}

\author{J{\"u}rgen P. Kropp}
\affiliation{Potsdam Institute for Climate Impact Research -- 
14412 Potsdam, Germany, EU}
\affiliation{
University of Potsdam, Dept. of Geo- \& Environmental Sciences, 
14476 Potsdam, Germany}

\date{\today, \jobname}

\begin{abstract}
In order to assess future damage caused by natural disasters, 
it is desirable to estimate the damage caused by single events. 
So called damage functions provide -- for a natural disaster of certain 
magnitude -- a specific damage value. 
However, in general, the functional form of such damage functions is unknown. 
We study the distributions of recorded flood damages on extended scales 
and deduce which damage functions lead to such distributions when the 
floods obey Generalized Extreme Value statistics 
and follow Generalized Pareto distributions. 
Based on the finding of broad damage distributions we investigate two possible 
functional forms to characterize the data.
In the case of Gumbel distributed extreme events, 
(i) a power-law distribution density with an exponent 
close to~$2$ (Zipf's law) implies an exponential damage function;
(ii) stretched exponential distribution densities 
imply power-law damage functions.
In the case of Weibull (Fr\'echet) distributed extreme events we find 
correspondingly steeper (less steep) damage functions. 
\end{abstract}

\maketitle

\section{Introduction}

Natural disasters, such as floods or storms, represent extreme events 
\cite{MasterOfDisaster2002,GadelHakM2008,KroppS2011,BundeEKH2005} 
with severe consequences including numerous killed and affected people 
as well as huge economic damage. 
Independent of the problem on how to project the occurrence of 
future extreme events, 
one is interested in which damage can typically be expected from an 
extreme event of certain magnitude. 
One approach to tackle this question is to separate the statistics of 
extreme events from the damage caused by them. 
The former can be obtained from measurements, such as water level records, 
but the latter needs to be estimated by some empirical studies.

Accordingly, so called damage functions provide a monetary value as a 
function of the magnitude of an event, such as the maximum flood level. 
We distinguish damage functions on a microscopic scale from those on a 
macroscopic scale \cite{MerzKST2010,BoettleKRRRW2011,PrahlRBK2016}. 
The former describes the typical costs of damages to single assets, 
such as residential buildings, and the latter describes damage costs at 
larger areas, such as an entire city.
This macroscopic (aggregated) damage function represents a 
composition of information on asset values, their location, 
and their vulnerability.
We assume that in this case the damages are well characterized 
by the function, i.e.\ noise is reduced due to spatial aggregation.

In general, the functional form of damage functions is unknown. 
By definition, they are monotonic increasing with the magnitude of the 
natural hazard and eventually exhibit saturation.
For floods, on the microscopic scale a considerable set of 
recorded direct monetary damage values of inundated buildings has been 
related to the corresponding water depth leading to an exponential 
dependence for private housing \cite[Fig.~5]{MerzKTS2004}.
Furthermore, a set of different microscopic damage functions, models, 
and the related damages have been compared and evaluated 
\cite{ApelAKT2009}, 
among them linear, quadratic, and square-root damage functions 
(see also \cite{SmithDI1994}). 
In \cite{HallegatteRMDCMHW2010} a storm surge damage function has been 
estimated for the city of Copenhagen.
In what follows we refer to macroscopic damage functions, 
i.e.\ the direct damage to an urban agglomeration as a function of 
the maximum flood height.

Since from single sites there are usually too few damage records 
to obtain a damage function, 
we tackle the problem on a larger scale and study
which functional form a 
damage function must follow so that the distribution of 
extreme events transforms to the distribution of observed damages. 
Relating these two distributions we obtain macroscopic damage
functions.
Therefore, we analyze flood data assembled by CRED \cite{CREDref} 
and find broad damage distributions. 
In order to characterize them, we elaborate two functional forms. 
We show that for the Gumbel case, 
Zipf's law, i.e.\ a power-law distribution density with 
an exponent~$\alpha\approx 2$, implies an exponential damage function.
Stretched exponential distribution densities imply 
power-law damage functions.

As it is known analytically,
maximum values of samples of fixed size 
(such as "block maxima" of time series)
follow distributions which converge 
for sufficiently large samples 
towards Generalized Extreme Value (GEV) distributions, 
if the values are independent and identically distributed.
Thus, we assume that changes in the statistics are small compared 
to the implied damages.
GEV distributions are commonly fitted to maxima 
in order to estimate annualities and future occurrences 
\cite{HawkesGSP2008}.

Further, if the data obeys GEV characteristics, 
it is possible to use the Generalized Pareto (GP) distributions
as an approximation of the distribution function of the level~$s$ 
above a sufficiently high threshold $s_T$ \cite{ColesS2001}.
Therefore, the upper tail of the distribution function of the 
(damage causing) events is described by one of the three GP distributions
\begin{equation}
  P_{(s)}^{\rm GP} = \left \{
    \begin{array}{ll}
      1 - \left ( 1 + \xi \dfrac{s - s_T}{\widetilde \gamma} \right )^{-\frac{1}{\xi}} & \mbox{for } \xi \neq 0 \\
      1 - {\rm e}^{-\frac{s - s_T}{\widetilde \gamma}} & \mbox{for } \xi = 0 \, .
    \end{array} \right .
    \label{eq:gpd}
\end{equation}
They are defined on $s \in [s_T,\infty)$ and have 
a scale parameter, $\widetilde \gamma \in \mathbb{R}^{+}$, as well as 
a shape parameter, $\xi \in \mathbb{R}$
(for $\xi<0$: $s \in (s_T,s_T-\frac{\widetilde\gamma}{\xi} )$).
According to the GEV-shape, one distinguishes three cases:
(i) the Gumbel distribution ($\xi = 0$), 
(ii) the heavy-tailed Fr\'echet distribution ($\xi > 0$), and 
(iii) the bounded-tailed \emph{reversed} Weibull distribution ($\xi < 0$).
For a more detailed presentation of extreme value assessment and applications 
we refer to \cite{ColesS2001,LeadbetterLR1983,KatzPN2002,EichnerKBH2006}.

\begin{figure}[ht]
\begin{center}
\includegraphics[width=\columnwidth]{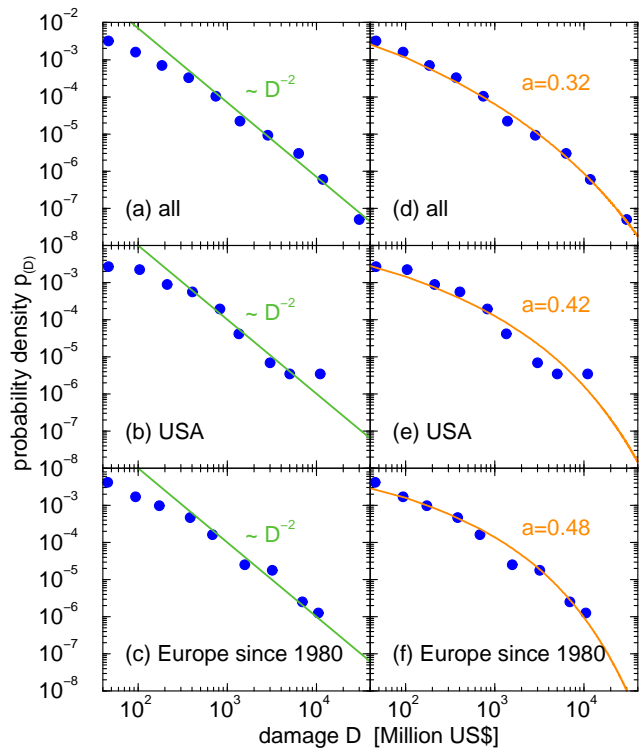}
\end{center}
\caption{
Probability densities of damages due to floods worldwide in the years 1950-2008 
as estimated from the records provided by CRED \cite{CREDref}.
(a+d) all entries, all years;
(b+e) USA only, all years, 70 events; and 
(c+f) Europe only, 1980-2008, 193 events.
The straight lines in (a-c) are guides to the eye corresponding to 
Eq.~(\ref{eq:power-law}) with $\alpha=2$.
The solid lines in (d-f) are stretched exponential fits according 
to Eq.~(\ref{eq:stretched}) providing the exponents 
(d) $a\simeq 0.32$,
(e) $a\simeq 0.42$, and 
(f) $a\simeq 0.48$.
The probability densities have been estimated in logarithmic bins.
}
\label{fig:damage}
\end{figure}

\section{Damage records}

We consider the EM-DAT database \cite{CREDref}
collected by the Centre for Research on Epidemiology of Disasters (CRED) 
in version v12.07 as created on Oct-28-2009. 
The information listed for each event entry consists of: 
start, end, country, location, type of disaster, sub-type, name, 
number of people killed, number of people affected, an estimated damage, 
and an ID. 
For the years 1950-2008 we extract the information on floods, 
which include general floods, flash floods, as well as storm surges 
respectively coastal floods, and obtain $3469$~entries worldwide, 
while for $1225$~entries an estimated damage in units of Million US-Dollars 
is available.

In Figure~\ref{fig:damage}(a) we show the estimated probability densities, 
$\widetilde p(D)$, for all flood damage values of the database. 
However, since one may argue that the result could be biased by regional 
differences, in Fig.~\ref{fig:damage}(b) we show $\widetilde p(D)$ 
for only those floods that occurred in the USA. 
Thus, we can weaken influences due to different
economic power of different countries.
In order to reduce possible trends in the data \cite{PielkeGLCSM2008}, 
we also exclude floods before 1980 and plot the distribution density for 
Europe in Fig.~\ref{fig:damage}(c). 
In any case, we observe broad distributions with damages reaching 
the order of $10$\,Billion US-Dollars.
We would also like to note that we obtain similar distributions 
for the number of killed or affected people as well as for other 
natural disasters.

In the simplest approach, the tail of the probability densities 
can be described with a functional form involving one parameter, 
namely a power-law according to
\begin{equation}
  \widetilde p_{(D)} \sim D^{-\alpha} \, ,
  \label{eq:power-law}
\end{equation}
where we find~$\alpha\approx 2$ [Fig.~\ref{fig:damage}(a-c)]. 
Such a size distribution is also known as Zipf's law 
(a special case of the Pareto distribution)
and is found in many different fields, 
such as word usage, city sizes, firm sizes, 
wealth, intensity of solar flares, etc.
For an overview we refer to \cite{NewmanMEJ2005,RozenfeldRGM2011,RybskiD2013} 
and references therein.
Minor deviations from Eq.~(\ref{eq:power-law}) for floods with small damage 
could be due to the fact that small damages are more likely to be 
missing in the database.

\section{Relating extreme events and damages}

The damage costs of a large flood magnitude event depend on a variety of 
damage influencing factors, 
such as orography, flow velocity, contamination, preparedness, 
etc.\ \cite{ThiekenMKM2005}.
Nevertheless, in the common practice, the dominant correlations 
with the maximum flood level are explored \cite{MerzKST2010}. 
Thus, if we assume a unique relation between the magnitude of 
an extreme event, $s$, and the mean damage, $D_{(s)}$, that is caused by it, 
we can take advantage of Eq.~(\ref{eq:gpd}) and~(\ref{eq:power-law}) 
and relate them. 
Therefore, we write the probability as an integral over the density and 
substitute the magnitude with the damage 
$\left( s \rightarrow D = D_{(s)} \right)$:
\begin{equation}
  \int \limits _{s_1} ^{s_2} p _{(s)} \mbox{d}s = \int \limits _{D_{(s_1)}} ^{D_{(s_2)}} p_{(s_{(D)})} \dfrac{\mbox{d}s}{\mbox{d}D} \mbox{d}D = \int \limits _{D_{(s_1)}} ^{D_{(s_2)}} \widetilde p_{(D)} \mbox{d}D \,.
  \label{eq:1}
\end{equation}
Here, $p_{(s)}$ and $\widetilde p_{(D)}$ are the probability density of the 
extreme event and the damage, respectively. 
Furthermore, the density transformation 
$\dfrac{\widetilde p _{(D)}}{p _{(s)}} = \dfrac{\mbox{d}s}{\mbox{d}D}$ 
was used.
Next, choosing~$s_1 = s_T$ 
(the threshold for which GP distributions are applicable) 
and~$s_2 = s$, we obtain the equation 
\begin{equation}
  P_{(s)} = \int \limits _{D_{(s_T)}} ^{D_{(s)}} \widetilde p_{(D)} \mbox{d}D 
  \, ,
  \label{eq:prob}
\end{equation}
which holds for any (reasonable) 
damage distribution density~$\widetilde p_{(D)}$.

Using Eq.~(\ref{eq:power-law}), i.e.\ 
the probability density~$\widetilde p_{(D)} = A D^{-\alpha}$ 
$(\alpha > 1)$ as indicated in Fig.~\ref{fig:damage}(a-c), we write 
\begin{equation}
  P_{(s)} = \int \limits_{D_{(s_T)}}^{D_{(s)}} A D^{-\alpha} \mbox{d}D
	  = 1 - D_{(s_T)} ^{\alpha-1} D_{(s)} ^{1-\alpha} \, ,
  \label{eq:prob-zipf}
\end{equation}
where $A=(\alpha-1) D_{(s_T)}^{\alpha-1}$ such that~$\widetilde p _{(D)}$ 
is normalized in~$D \in [D_{(s_T)},D_{(s\rightarrow \infty)})$. 
Further we suppose that~$ D_{(s \rightarrow \infty)} \rightarrow \infty$. 
Solving the equation for~$D_{(s)}$ we get
\begin{equation}
  D_{(s)} = D_{(s_T)} \left( 1-P_{(s)} \right)^{\frac{1}{1-\alpha}} \, . 
  \label{eq:solution}
\end{equation}
Finally, we insert the GP distribution for the Gumbel case, 
i.e.\ Eq.~(\ref{eq:gpd}) with~$\xi=0$, and obtain that the damage
increases exponentially,
\begin{equation}
  D_{(s)} = D_{(s_T)} {\rm e}^{\frac{s - s_T}{\widetilde \gamma (\alpha - 1)}} 
  \, ,
  \label{eq:gpd-solved}
\end{equation}
which holds for $s \in [s_T,\infty)$.
Accordingly, the exponential damage function transforms 
the Gumbel distribution, approximated by an exponential one, 
into a Pareto distribution \cite{ReissT2007Ch1}. 

However, the exponential damage function is based on damages following 
power-law distributions [Fig.~\ref{fig:damage}(a-c)].
Since a one parameter description might not be sufficient, we also elaborate 
a two parameter fit -- which we choose since it is integrable -- 
namely a stretched exponential according to
\begin{equation}
  \widetilde p(D) \sim \frac{a}{b} D^{a-1} {\rm e}^{-\frac{D^{a}}{b}} 
  \, ,
  \label{eq:stretched}
\end{equation}
where~$a$ and~$b$ are the parameters ($a,b>0$). 
Equation~(\ref{eq:stretched}) is also known as Weibull distribution, 
see e.g.\ \cite{LaherrereS1998,IvanovYPL2004} and references therein.
For the same data as before the fitted curves are shown in 
Fig.~\ref{fig:damage}(d-f) providing values for the exponent~$a$ 
roughly between~$1/3$ and~$1/2$.
The stretched exponential (Weibull) distribution, 
Eq.~(\ref{eq:stretched}), is sometimes used as penultimate approximation 
for pre-asymptotic behavior of extreme value distributions 
\cite{WilsonT2005,FurrerK2008}.
This suggests, that the usage of the stretched exponential distribution 
could be justified by an insufficiently high damage threshold.

Now, the integral relating the extreme events~$s$ and damages~$D$, 
Eq.~(\ref{eq:prob}), is over $\widetilde p(D)$ 
from Eq.~(\ref{eq:stretched}), instead of Eq.~(\ref{eq:power-law}):
\begin{equation}
   P_{(s)} = \int\limits_{D_{(s_T)}}^{D_{(s)}} \!\! B\frac{a}{b} D^{a-1} {\rm e}^{-\frac{D^{a}}{b}} \mbox{d}D
           = 1 - B \, {\rm e}^{-\frac{D_{(s)}^{a}}{b}} \, ,
  \label{eq:prob-stretch}
\end{equation}
where $B={\rm e}^{\frac{D_{(s_T)}^{a}}{b}}$ 
such that~$\widetilde p _{(D)}$ is 
normalized in~$D \in [D_{(s_T)},D_{(s\rightarrow \infty)})$.
Solving for $D_{(s)}$ we find
\begin{equation}
  D_{(s)} = \left[D_{(s_T)}^{a}-b~\ln(1-P_{(s)})\right]^{\frac{1}{a}} 
  \, .
  \label{eq:DsPsSE}
\end{equation}
Finally, we insert the GP distribution for the Gumbel case, 
i.e.\ Eq.~(\ref{eq:gpd}) with $\xi=0$, 
and obtain 
\begin{equation}
  D_{(s)} = \left( D^{a}_{(s_T)} + \frac{b}{\widetilde \gamma} (s-s_T) \right)^{\frac{1}{a}}
  \, .
  \label{eq:GEV-strech}
\end{equation}
Accordingly, the power-law damage function transforms 
the Gumbel distribution, approximated by an exponential one, 
into a stretched exponential distribution \cite{JohnsonKB1994Ch20}.
For~$a$ between~$1/3$ and~$1/2$ 
we obtain the asymptotic power-law relation~$D_{(s)} \sim s^{3}$ 
or~$D_{(s)} \sim s^{2}$, respectively.

\section{Summary and Discussion}
\label{sec:conclude}

\begin{figure*}[t]
\begin{center}
\includegraphics[width=0.75\textwidth]{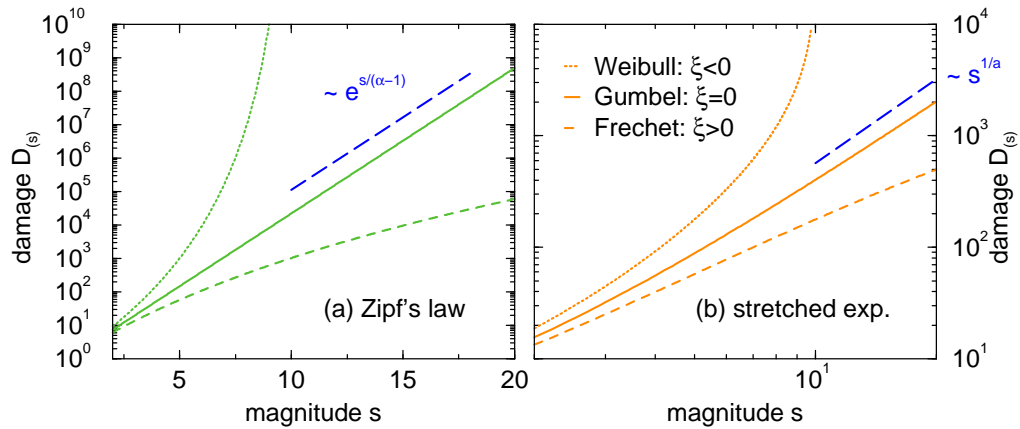}
\end{center}
\caption{
Illustration of the obtained damage functions.
(a) In the case of power-law distributed damages for $\xi=0$ we obtain an 
exponential damage function [Eq.~(\ref{eq:gpd-solved}), solid line].
For $\xi\ne0$ we insert Eq.~(\ref{eq:gpd}) in Eq.~(\ref{eq:solution}) and find 
faster (dotted) or slower (dashed) than exponential damage functions, 
Weibull ($\xi<0$) or Fr\'echet ($\xi>0$) case, respectively.
(b) In the case of stretched exponential distributed damages for $\xi=0$ 
we obtain a power-law damage function with exponent~$1/a$
[Eq.~(\ref{eq:GEV-strech}), solid line]. 
For $\xi\ne0$ we insert Eq.~(\ref{eq:gpd}) in Eq.~(\ref{eq:DsPsSE}) 
and find that the damage function increases faster (dotted) or slower (dashed), 
Weibull ($\xi<0$) or Fr\'echet ($\xi>0$) case, respectively, 
than the corresponding power-law.
To generate the curves, we choose: 
Fr\'echet,~$\xi=0.1$ and Weibull,~$\xi=-0.1$ 
as well as $\alpha=2$ in (a) and $a=0.4$ in (b).
}
\label{fig:resillu}
\end{figure*}

In summary, we characterize distributions of recorded flood damages, 
argue that they are caused by extreme events, and employ 
density transformation to deduce damage functions relating both.
For Gumbel distributed extreme events ($\xi=0$) we find asymptotically 
\begin{equation}
  D_{(s)} \sim 
    \left\{
      \begin{array}{ll}
        {\rm e}^\frac{s}{\widetilde\gamma(\alpha-1)} & \mbox{for } \widetilde p_{(D)} \sim D^{-\alpha} 
	  \mbox{ with } \alpha>1
	  \vspace{0.1cm}
	  \\
	\left(\frac{1}{\widetilde\gamma}s\right)^\frac{1}{a} & \mbox{for } \widetilde p_{(D)} \sim \frac{a}{b} D^{a-1} {\rm e}^{-\frac{D^{a}}{b}} 
	  \mbox{ with } a>0
	  \, . \\
      \end{array}
    \right .
  \label{eq:summary}
\end{equation}
The involved GP parameters are in an ultimate sense and in practice 
penultimate approximations might be necessary \cite{FurrerK2008,WilsonT2005}. 
In particular, it needs to be considered that if the 
underlying geophysical variable has only an approximate exponential 
distribution [i.e.\ $\xi=0$ in Eq.~(\ref{eq:gpd})], then the form of 
the obtained damage functions would not be unique.
An analogous argument applies to the cases $\xi\ne 0$ and 
further research is necessary to find unique relations.

The functional forms of Eq.~(\ref{eq:summary}) 
are illustrated in Fig.~\ref{fig:resillu}, which also 
includes the cases $\xi\ne 0$. 
For power-law distributed damages and Weibull distributed 
extreme events ($\xi<0$) in average the damage increases faster than 
exponentially with the magnitude. 
For Fr\'echet distributed extreme events ($\xi>0$) the opposite is the case.
Intuitively, since in the Weibull case the extreme events have a 
less heavy tail, a steeper damage function is needed to result in the same 
damage distribution as the Gumbel case [Fig.~\ref{fig:resillu}(a)]. 
In the Fr\'echet case, which has a fatter tail than the Gumbel distribution, 
a less steep damage function is sufficient to result in the same 
damage distribution as the Gumbel case.
Similar arguments hold for stretched exponential damages 
[Fig.~\ref{fig:resillu}(b)].

We would like to remark that, 
allowing also negative values of~$a$ in Eq.~(\ref{eq:stretched}), 
it represents the Fr\'echet instead of the Weibull distribution. 
Moreover, with a scale parameter~$a=-1$, in the limit 
$s \rightarrow \infty$ the Fr\'echet distribution 
is equivalent to Zipf's law, Eq.~(\ref{eq:power-law}).

In particular the role of aggregation on both, the extremes and 
the damages functions requires further investigations.
While floods involve an integration of precipitation over a water basin, 
damages involve an integration of flood impacts over affected assets. 
Thus, it would be interesting to understand, to which extend the 
tail becomes heavier at each stage, from precipitation extremes to
flood magnitude and then to damages.

Since the obtained damage functions represent an average for large 
temporal and spatial scale, neglecting any further local differences, 
they have limited predictive power. 
However, the results could provide qualitative insight because exponential 
increasing damage is much more catastrophic than polynomial. 
Thus, it would be of interest to systematically analyze how the results 
depend on the considered spatial scale.

Our approach could help to address the questions raised by 
decision makers or insurances, which costs certain regions, countries, 
or even the globalized economy are facing from 
coastal floods \cite{BoettleRK2013,BoettleRK2015,BoettleRK2016}.
The assessment of damage costs is fundamental in the econtext of 
adaptation to climate change \cite{SchmidtThomeK2013}.
Finally, we would like to note that the presented approach can be applied to 
any type of natural disasters as long as the concept of damage functions 
can be used (see e.g.\ \cite{PrahlRKBH2012,PrahlRBK2015}).

\begin{acknowledgments}
We would like to thank M.~Boettle for useful discussions and 
several anonymous reviewers for helpful comments.
We appreciate financial support from BaltCICA 
(part-financed by the EU Baltic Sea Region Programme 2007-2013) 
and from the Cham\"aleon-Project 
(financed by the German Federal Ministry of Education and Research).
\end{acknowledgments}


\begin{thebibliography}{34}%
\makeatletter
\providecommand \@ifxundefined [1]{%
 \@ifx{#1\undefined}
}%
\providecommand \@ifnum [1]{%
 \ifnum #1\expandafter \@firstoftwo
 \else \expandafter \@secondoftwo
 \fi
}%
\providecommand \@ifx [1]{%
 \ifx #1\expandafter \@firstoftwo
 \else \expandafter \@secondoftwo
 \fi
}%
\providecommand \natexlab [1]{#1}%
\providecommand \enquote  [1]{``#1''}%
\providecommand \bibnamefont  [1]{#1}%
\providecommand \bibfnamefont [1]{#1}%
\providecommand \citenamefont [1]{#1}%
\providecommand \href@noop [0]{\@secondoftwo}%
\providecommand \href [0]{\begingroup \@sanitize@url \@href}%
\providecommand \@href[1]{\@@startlink{#1}\@@href}%
\providecommand \@@href[1]{\endgroup#1\@@endlink}%
\providecommand \@sanitize@url [0]{\catcode `\\12\catcode `\$12\catcode
  `\&12\catcode `\#12\catcode `\^12\catcode `\_12\catcode `\%12\relax}%
\providecommand \@@startlink[1]{}%
\providecommand \@@endlink[0]{}%
\providecommand \url  [0]{\begingroup\@sanitize@url \@url }%
\providecommand \@url [1]{\endgroup\@href {#1}{\urlprefix }}%
\providecommand \urlprefix  [0]{URL }%
\providecommand \Eprint [0]{\href }%
\providecommand \doibase [0]{http://dx.doi.org/}%
\providecommand \selectlanguage [0]{\@gobble}%
\providecommand \bibinfo  [0]{\@secondoftwo}%
\providecommand \bibfield  [0]{\@secondoftwo}%
\providecommand \translation [1]{[#1]}%
\providecommand \BibitemOpen [0]{}%
\providecommand \bibitemStop [0]{}%
\providecommand \bibitemNoStop [0]{.\EOS\space}%
\providecommand \EOS [0]{\spacefactor3000\relax}%
\providecommand \BibitemShut  [1]{\csname bibitem#1\endcsname}%
\let\auto@bib@innerbib\@empty
\bibitem [{\citenamefont {Bunde}\ \emph {et~al.}(2002)\citenamefont {Bunde},
  \citenamefont {Kropp},\ and\ \citenamefont
  {Schellnhuber}}]{MasterOfDisaster2002}%
  \BibitemOpen
  \bibinfo {editor} {\bibfnamefont {A.}~\bibnamefont {Bunde}}, \bibinfo
  {editor} {\bibfnamefont {J.}~\bibnamefont {Kropp}}, \ and\ \bibinfo {editor}
  {\bibfnamefont {H.-J.}\ \bibnamefont {Schellnhuber}},\ eds.,\ \href@noop {}
  {\emph {\bibinfo {title} {The Science of Disasters -- Climate Disruptions,
  Heart Attacks, and Market Crashes}}}\ (\bibinfo  {publisher}
  {Springer-Verlag},\ \bibinfo {address} {Berlin},\ \bibinfo {year}
  {2002})\BibitemShut {NoStop}%
\bibitem [{\citenamefont {Gad-el Hak}(2008)}]{GadelHakM2008}%
  \BibitemOpen
  \bibinfo {editor} {\bibfnamefont {M.}~\bibnamefont {Gad-el Hak}},\ ed.,\
  \href@noop {} {\emph {\bibinfo {title} {Large-scale disasters -- prediction,
  control, and mitigation}}}\ (\bibinfo  {publisher} {Cambridge University
  Press},\ \bibinfo {address} {Cambridge},\ \bibinfo {year} {2008})\BibitemShut
  {NoStop}%
\bibitem [{\citenamefont {Kropp}\ and\ \citenamefont
  {Schellnhuber}(2011)}]{KroppS2011}%
  \BibitemOpen
  \bibinfo {editor} {\bibfnamefont {J.}~\bibnamefont {Kropp}}\ and\ \bibinfo
  {editor} {\bibfnamefont {H.-J.}\ \bibnamefont {Schellnhuber}},\ eds.,\
  \href@noop {} {\emph {\bibinfo {title} {In Extremis -- Disruptive Events and
  Trends in Climate and Hydrology}}}\ (\bibinfo  {publisher}
  {Springer-Verlag},\ \bibinfo {address} {Berlin},\ \bibinfo {year}
  {2011})\BibitemShut {NoStop}%
\bibitem [{\citenamefont {Bunde}\ \emph {et~al.}(2005)\citenamefont {Bunde},
  \citenamefont {Eichner}, \citenamefont {Kantelhardt},\ and\ \citenamefont
  {Havlin}}]{BundeEKH2005}%
  \BibitemOpen
  \bibfield  {author} {\bibinfo {author} {\bibfnamefont {A.}~\bibnamefont
  {Bunde}}, \bibinfo {author} {\bibfnamefont {J.~F.}\ \bibnamefont {Eichner}},
  \bibinfo {author} {\bibfnamefont {J.~W.}\ \bibnamefont {Kantelhardt}}, \ and\
  \bibinfo {author} {\bibfnamefont {S.}~\bibnamefont {Havlin}},\ }\href@noop {}
  {\bibfield  {journal} {\bibinfo  {journal} {Phys. Rev. Lett.}\ }\textbf
  {\bibinfo {volume} {94}},\ \bibinfo {pages} {048701} (\bibinfo {year}
  {2005})}\BibitemShut {NoStop}%
\bibitem [{\citenamefont {Merz}\ \emph {et~al.}(2010)\citenamefont {Merz},
  \citenamefont {Kreibich}, \citenamefont {Schwarze},\ and\ \citenamefont
  {Thieken}}]{MerzKST2010}%
  \BibitemOpen
  \bibfield  {author} {\bibinfo {author} {\bibfnamefont {B.}~\bibnamefont
  {Merz}}, \bibinfo {author} {\bibfnamefont {H.}~\bibnamefont {Kreibich}},
  \bibinfo {author} {\bibfnamefont {R.}~\bibnamefont {Schwarze}}, \ and\
  \bibinfo {author} {\bibfnamefont {A.~H.}\ \bibnamefont {Thieken}},\
  }\href@noop {} {\bibfield  {journal} {\bibinfo  {journal} {Nat Hazard Earth
  Sys}\ }\textbf {\bibinfo {volume} {10}},\ \bibinfo {pages} {1697} (\bibinfo
  {year} {2010})}\BibitemShut {NoStop}%
\bibitem [{\citenamefont {Boettle}\ \emph {et~al.}(2011)\citenamefont
  {Boettle}, \citenamefont {Kropp}, \citenamefont {Reiber}, \citenamefont
  {Roithmeier}, \citenamefont {Rybski},\ and\ \citenamefont
  {Walther}}]{BoettleKRRRW2011}%
  \BibitemOpen
  \bibfield  {author} {\bibinfo {author} {\bibfnamefont {M.}~\bibnamefont
  {Boettle}}, \bibinfo {author} {\bibfnamefont {J.~P.}\ \bibnamefont {Kropp}},
  \bibinfo {author} {\bibfnamefont {L.}~\bibnamefont {Reiber}}, \bibinfo
  {author} {\bibfnamefont {O.}~\bibnamefont {Roithmeier}}, \bibinfo {author}
  {\bibfnamefont {D.}~\bibnamefont {Rybski}}, \ and\ \bibinfo {author}
  {\bibfnamefont {C.}~\bibnamefont {Walther}},\ }\href@noop {} {\bibfield
  {journal} {\bibinfo  {journal} {Nat. Hazards Earth Syst. Sci.}\ }\textbf
  {\bibinfo {volume} {11}},\ \bibinfo {pages} {3327} (\bibinfo {year}
  {2011})}\BibitemShut {NoStop}%
\bibitem [{\citenamefont {Prahl}\ \emph {et~al.}(2016)\citenamefont {Prahl},
  \citenamefont {Rybski}, \citenamefont {Boettle},\ and\ \citenamefont
  {Kropp}}]{PrahlRBK2016}%
  \BibitemOpen
  \bibfield  {author} {\bibinfo {author} {\bibfnamefont {B.~F.}\ \bibnamefont
  {Prahl}}, \bibinfo {author} {\bibfnamefont {D.}~\bibnamefont {Rybski}},
  \bibinfo {author} {\bibfnamefont {M.}~\bibnamefont {Boettle}}, \ and\
  \bibinfo {author} {\bibfnamefont {J.~P.}\ \bibnamefont {Kropp}},\ }\href@noop
  {} {\bibfield  {journal} {\bibinfo  {journal} {in prep.}\ } (\bibinfo {year}
  {2016})}\BibitemShut {NoStop}%
\bibitem [{\citenamefont {Merz}\ \emph {et~al.}(2004)\citenamefont {Merz},
  \citenamefont {Kreibich}, \citenamefont {Thieken},\ and\ \citenamefont
  {Schmidtke}}]{MerzKTS2004}%
  \BibitemOpen
  \bibfield  {author} {\bibinfo {author} {\bibfnamefont {B.}~\bibnamefont
  {Merz}}, \bibinfo {author} {\bibfnamefont {H.}~\bibnamefont {Kreibich}},
  \bibinfo {author} {\bibfnamefont {A.}~\bibnamefont {Thieken}}, \ and\
  \bibinfo {author} {\bibfnamefont {R.}~\bibnamefont {Schmidtke}},\ }\href@noop
  {} {\bibfield  {journal} {\bibinfo  {journal} {Nat. Hazards Earth Syst.
  Sci.}\ }\textbf {\bibinfo {volume} {4}},\ \bibinfo {pages} {153} (\bibinfo
  {year} {2004})}\BibitemShut {NoStop}%
\bibitem [{\citenamefont {Apel}\ \emph {et~al.}(2009)\citenamefont {Apel},
  \citenamefont {Aronica}, \citenamefont {Kreibich},\ and\ \citenamefont
  {Thieken}}]{ApelAKT2009}%
  \BibitemOpen
  \bibfield  {author} {\bibinfo {author} {\bibfnamefont {H.}~\bibnamefont
  {Apel}}, \bibinfo {author} {\bibfnamefont {G.~T.}\ \bibnamefont {Aronica}},
  \bibinfo {author} {\bibfnamefont {H.}~\bibnamefont {Kreibich}}, \ and\
  \bibinfo {author} {\bibfnamefont {A.~H.}\ \bibnamefont {Thieken}},\
  }\href@noop {} {\bibfield  {journal} {\bibinfo  {journal} {Nat. Hazards}\
  }\textbf {\bibinfo {volume} {49}},\ \bibinfo {pages} {79} (\bibinfo {year}
  {2009})}\BibitemShut {NoStop}%
\bibitem [{\citenamefont {Smith}(1994)}]{SmithDI1994}%
  \BibitemOpen
  \bibfield  {author} {\bibinfo {author} {\bibfnamefont {D.~I.}\ \bibnamefont
  {Smith}},\ }\href@noop {} {\bibfield  {journal} {\bibinfo  {journal} {Water
  SA}\ }\textbf {\bibinfo {volume} {20}},\ \bibinfo {pages} {231} (\bibinfo
  {year} {1994})}\BibitemShut {NoStop}%
\bibitem [{\citenamefont {Hallegatte}\ \emph {et~al.}(2010)\citenamefont
  {Hallegatte}, \citenamefont {Ranger}, \citenamefont {Mestre}, \citenamefont
  {Dumas}, \citenamefont {Corfee-Morlot}, \citenamefont {Herweijer},\ and\
  \citenamefont {Wood}}]{HallegatteRMDCMHW2010}%
  \BibitemOpen
  \bibfield  {author} {\bibinfo {author} {\bibfnamefont {S.}~\bibnamefont
  {Hallegatte}}, \bibinfo {author} {\bibfnamefont {N.}~\bibnamefont {Ranger}},
  \bibinfo {author} {\bibfnamefont {O.}~\bibnamefont {Mestre}}, \bibinfo
  {author} {\bibfnamefont {P.}~\bibnamefont {Dumas}}, \bibinfo {author}
  {\bibfnamefont {J.}~\bibnamefont {Corfee-Morlot}}, \bibinfo {author}
  {\bibfnamefont {C.}~\bibnamefont {Herweijer}}, \ and\ \bibinfo {author}
  {\bibfnamefont {R.~M.}\ \bibnamefont {Wood}},\ }\href@noop {} {\bibfield
  {journal} {\bibinfo  {journal} {Clim. Change}\ }\textbf {\bibinfo {volume}
  {104}},\ \bibinfo {pages} {113} (\bibinfo {year} {2010})}\BibitemShut
  {NoStop}%
\bibitem [{\citenamefont {EM-DAT}(2009)}]{CREDref}%
  \BibitemOpen
  \bibfield  {author} {\bibinfo {author} {\bibnamefont {EM-DAT}},\ }\href@noop
  {} {} (\bibinfo {year} {2009}),\ \bibinfo {note} {{EM-DAT}: The OFDA/CRED
  International Disaster Database - www.emdat.be, Universit\'e Catholique de
  Louvain, Brussels (Belgium)}\BibitemShut {NoStop}%
\bibitem [{\citenamefont {Hawkes}\ \emph {et~al.}(2008)\citenamefont {Hawkes},
  \citenamefont {Gonzalez-Marco}, \citenamefont {Sanchez-Arcilla},\ and\
  \citenamefont {Prinos}}]{HawkesGSP2008}%
  \BibitemOpen
  \bibfield  {author} {\bibinfo {author} {\bibfnamefont {P.~J.}\ \bibnamefont
  {Hawkes}}, \bibinfo {author} {\bibfnamefont {D.}~\bibnamefont
  {Gonzalez-Marco}}, \bibinfo {author} {\bibfnamefont {A.}~\bibnamefont
  {Sanchez-Arcilla}}, \ and\ \bibinfo {author} {\bibfnamefont {P.}~\bibnamefont
  {Prinos}},\ }\href@noop {} {\bibfield  {journal} {\bibinfo  {journal} {J.
  Hydraul. Res.}\ }\textbf {\bibinfo {volume} {46}},\ \bibinfo {pages} {323}
  (\bibinfo {year} {2008})}\BibitemShut {NoStop}%
\bibitem [{\citenamefont {Coles}(2001)}]{ColesS2001}%
  \BibitemOpen
  \bibfield  {author} {\bibinfo {author} {\bibfnamefont {S.}~\bibnamefont
  {Coles}},\ }\href@noop {} {\emph {\bibinfo {title} {An Introduction to
  Statistical Modeling of Extreme Values}}},\ Springer Series in Statistics\
  (\bibinfo  {publisher} {Springer},\ \bibinfo {address} {London},\ \bibinfo
  {year} {2001})\BibitemShut {NoStop}%
\bibitem [{\citenamefont {Leadbetter}\ \emph {et~al.}(1983)\citenamefont
  {Leadbetter}, \citenamefont {Lindgren},\ and\ \citenamefont
  {Rootzen}}]{LeadbetterLR1983}%
  \BibitemOpen
  \bibfield  {author} {\bibinfo {author} {\bibfnamefont {M.~R.}\ \bibnamefont
  {Leadbetter}}, \bibinfo {author} {\bibfnamefont {G.}~\bibnamefont
  {Lindgren}}, \ and\ \bibinfo {author} {\bibfnamefont {H.}~\bibnamefont
  {Rootzen}},\ }\href@noop {} {\emph {\bibinfo {title} {Extremes and Related
  Properties of Random Sequences and Processes}}},\ Springer Series in
  Statistics\ (\bibinfo  {publisher} {Springer},\ \bibinfo {address} {New
  York},\ \bibinfo {year} {1983})\BibitemShut {NoStop}%
\bibitem [{\citenamefont {Katz}\ \emph {et~al.}(2002)\citenamefont {Katz},
  \citenamefont {Parlange},\ and\ \citenamefont {Naveau}}]{KatzPN2002}%
  \BibitemOpen
  \bibfield  {author} {\bibinfo {author} {\bibfnamefont {R.~W.}\ \bibnamefont
  {Katz}}, \bibinfo {author} {\bibfnamefont {M.~B.}\ \bibnamefont {Parlange}},
  \ and\ \bibinfo {author} {\bibfnamefont {P.}~\bibnamefont {Naveau}},\
  }\href@noop {} {\bibfield  {journal} {\bibinfo  {journal} {Adv. Water
  Resour.}\ }\textbf {\bibinfo {volume} {25}},\ \bibinfo {pages} {1287}
  (\bibinfo {year} {2002})}\BibitemShut {NoStop}%
\bibitem [{\citenamefont {Eichner}\ \emph {et~al.}(2006)\citenamefont
  {Eichner}, \citenamefont {Kantelhardt}, \citenamefont {Bunde},\ and\
  \citenamefont {Havlin}}]{EichnerKBH2006}%
  \BibitemOpen
  \bibfield  {author} {\bibinfo {author} {\bibfnamefont {J.~F.}\ \bibnamefont
  {Eichner}}, \bibinfo {author} {\bibfnamefont {J.~W.}\ \bibnamefont
  {Kantelhardt}}, \bibinfo {author} {\bibfnamefont {A.}~\bibnamefont {Bunde}},
  \ and\ \bibinfo {author} {\bibfnamefont {S.}~\bibnamefont {Havlin}},\
  }\href@noop {} {\bibfield  {journal} {\bibinfo  {journal} {Phys. Rev. E}\
  }\textbf {\bibinfo {volume} {73}},\ \bibinfo {pages} {016130} (\bibinfo
  {year} {2006})}\BibitemShut {NoStop}%
\bibitem [{\citenamefont {Pielke~Jr.}\ \emph {et~al.}(2008)\citenamefont
  {Pielke~Jr.}, \citenamefont {Gratz}, \citenamefont {Landsea}, \citenamefont
  {Collins}, \citenamefont {Saunders},\ and\ \citenamefont
  {Musulin}}]{PielkeGLCSM2008}%
  \BibitemOpen
  \bibfield  {author} {\bibinfo {author} {\bibfnamefont {R.~A.}\ \bibnamefont
  {Pielke~Jr.}}, \bibinfo {author} {\bibfnamefont {J.}~\bibnamefont {Gratz}},
  \bibinfo {author} {\bibfnamefont {C.~W.}\ \bibnamefont {Landsea}}, \bibinfo
  {author} {\bibfnamefont {D.}~\bibnamefont {Collins}}, \bibinfo {author}
  {\bibfnamefont {M.~A.}\ \bibnamefont {Saunders}}, \ and\ \bibinfo {author}
  {\bibfnamefont {R.}~\bibnamefont {Musulin}},\ }\href@noop {} {\bibfield
  {journal} {\bibinfo  {journal} {Nat. Haz. Rev.}\ }\textbf {\bibinfo {volume}
  {9}},\ \bibinfo {pages} {29} (\bibinfo {year} {2008})}\BibitemShut {NoStop}%
\bibitem [{\citenamefont {Newman}(2005)}]{NewmanMEJ2005}%
  \BibitemOpen
  \bibfield  {author} {\bibinfo {author} {\bibfnamefont {M.~E.~J.}\
  \bibnamefont {Newman}},\ }\href@noop {} {\bibfield  {journal} {\bibinfo
  {journal} {Contemp. Phys.}\ }\textbf {\bibinfo {volume} {46}},\ \bibinfo
  {pages} {323} (\bibinfo {year} {2005})}\BibitemShut {NoStop}%
\bibitem [{\citenamefont {Rozenfeld}\ \emph {et~al.}(2011)\citenamefont
  {Rozenfeld}, \citenamefont {Rybski}, \citenamefont {Gabaix},\ and\
  \citenamefont {Makse}}]{RozenfeldRGM2011}%
  \BibitemOpen
  \bibfield  {author} {\bibinfo {author} {\bibfnamefont {H.~D.}\ \bibnamefont
  {Rozenfeld}}, \bibinfo {author} {\bibfnamefont {D.}~\bibnamefont {Rybski}},
  \bibinfo {author} {\bibfnamefont {X.}~\bibnamefont {Gabaix}}, \ and\ \bibinfo
  {author} {\bibfnamefont {H.~A.}\ \bibnamefont {Makse}},\ }\href@noop {}
  {\bibfield  {journal} {\bibinfo  {journal} {Am. Econ. Rev.}\ ,\ \bibinfo
  {pages} {2205}} (\bibinfo {year} {2011})}\BibitemShut {NoStop}%
\bibitem [{\citenamefont {Rybski}(2013)}]{RybskiD2013}%
  \BibitemOpen
  \bibfield  {author} {\bibinfo {author} {\bibfnamefont {D.}~\bibnamefont
  {Rybski}},\ }\href {\doibase 10.1068/a4678} {\bibfield  {journal} {\bibinfo
  {journal} {Environ. Plan. A}\ }\textbf {\bibinfo {volume} {45}},\ \bibinfo
  {pages} {1266} (\bibinfo {year} {2013})}\BibitemShut {NoStop}%
\bibitem [{\citenamefont {Thieken}\ \emph {et~al.}(2005)\citenamefont
  {Thieken}, \citenamefont {M{\"u}ller}, \citenamefont {Kreibich},\ and\
  \citenamefont {Merz}}]{ThiekenMKM2005}%
  \BibitemOpen
  \bibfield  {author} {\bibinfo {author} {\bibfnamefont {A.~H.}\ \bibnamefont
  {Thieken}}, \bibinfo {author} {\bibfnamefont {M.}~\bibnamefont {M{\"u}ller}},
  \bibinfo {author} {\bibfnamefont {H.}~\bibnamefont {Kreibich}}, \ and\
  \bibinfo {author} {\bibfnamefont {B.}~\bibnamefont {Merz}},\ }\href@noop {}
  {\bibfield  {journal} {\bibinfo  {journal} {Water Resour. Res.}\ }\textbf
  {\bibinfo {volume} {41}} (\bibinfo {year} {2005})}\BibitemShut {NoStop}%
\bibitem [{\citenamefont {Reiss}\ and\ \citenamefont
  {Thomas}(2007)}]{ReissT2007Ch1}%
  \BibitemOpen
  \bibfield  {author} {\bibinfo {author} {\bibfnamefont {R.-D.}\ \bibnamefont
  {Reiss}}\ and\ \bibinfo {author} {\bibfnamefont {M.}~\bibnamefont {Thomas}},\
  }\enquote {\bibinfo {title} {Statistical analysis of extreme values: with
  applications to insurance, finance, hydrology and other fields},}\ \
  (\bibinfo  {publisher} {Birkh\"auser},\ \bibinfo {address} {Basel, Boston,
  Berlin},\ \bibinfo {year} {2007})\ Chap.\ \bibinfo {chapter} {1. Parametric
  Modeling}, pp.\ \bibinfo {pages} {3--38}\BibitemShut {NoStop}%
\bibitem [{\citenamefont {Laherrere}\ and\ \citenamefont
  {Sornette}(1998)}]{LaherrereS1998}%
  \BibitemOpen
  \bibfield  {author} {\bibinfo {author} {\bibfnamefont {J.}~\bibnamefont
  {Laherrere}}\ and\ \bibinfo {author} {\bibfnamefont {D.}~\bibnamefont
  {Sornette}},\ }\href@noop {} {\bibfield  {journal} {\bibinfo  {journal} {Eur.
  Phys. J. B}\ }\textbf {\bibinfo {volume} {2}},\ \bibinfo {pages} {525}
  (\bibinfo {year} {1998})}\BibitemShut {NoStop}%
\bibitem [{\citenamefont {Ivanov}\ \emph {et~al.}(2004)\citenamefont {Ivanov},
  \citenamefont {Yuen}, \citenamefont {Podobnik},\ and\ \citenamefont
  {Lee}}]{IvanovYPL2004}%
  \BibitemOpen
  \bibfield  {author} {\bibinfo {author} {\bibfnamefont {P.~C.}\ \bibnamefont
  {Ivanov}}, \bibinfo {author} {\bibfnamefont {A.}~\bibnamefont {Yuen}},
  \bibinfo {author} {\bibfnamefont {B.}~\bibnamefont {Podobnik}}, \ and\
  \bibinfo {author} {\bibfnamefont {Y.}~\bibnamefont {Lee}},\ }\href@noop {}
  {\bibfield  {journal} {\bibinfo  {journal} {Phys. Rev. E}\ }\textbf {\bibinfo
  {volume} {69}},\ \bibinfo {pages} {056107} (\bibinfo {year}
  {2004})}\BibitemShut {NoStop}%
\bibitem [{\citenamefont {Wilson}\ and\ \citenamefont
  {Toumi}(2005)}]{WilsonT2005}%
  \BibitemOpen
  \bibfield  {author} {\bibinfo {author} {\bibfnamefont {P.~S.}\ \bibnamefont
  {Wilson}}\ and\ \bibinfo {author} {\bibfnamefont {R.}~\bibnamefont {Toumi}},\
  }\href@noop {} {\bibfield  {journal} {\bibinfo  {journal} {Geophys. Res.
  Lett.}\ }\textbf {\bibinfo {volume} {32}},\ \bibinfo {pages} {L14812}
  (\bibinfo {year} {2005})}\BibitemShut {NoStop}%
\bibitem [{\citenamefont {Furrer}\ and\ \citenamefont
  {Katz}(2008)}]{FurrerK2008}%
  \BibitemOpen
  \bibfield  {author} {\bibinfo {author} {\bibfnamefont {E.~M.}\ \bibnamefont
  {Furrer}}\ and\ \bibinfo {author} {\bibfnamefont {R.~W.}\ \bibnamefont
  {Katz}},\ }\href@noop {} {\bibfield  {journal} {\bibinfo  {journal} {Water
  Resour. Res.}\ }\textbf {\bibinfo {volume} {44}},\ \bibinfo {pages} {W12439}
  (\bibinfo {year} {2008})}\BibitemShut {NoStop}%
\bibitem [{\citenamefont {Johnson}\ \emph {et~al.}(1994)\citenamefont
  {Johnson}, \citenamefont {Kotz},\ and\ \citenamefont
  {Balakrishnan}}]{JohnsonKB1994Ch20}%
  \BibitemOpen
  \bibfield  {author} {\bibinfo {author} {\bibfnamefont {N.~L.}\ \bibnamefont
  {Johnson}}, \bibinfo {author} {\bibfnamefont {S.}~\bibnamefont {Kotz}}, \
  and\ \bibinfo {author} {\bibfnamefont {N.}~\bibnamefont {Balakrishnan}},\
  }\enquote {\bibinfo {title} {Continuous univariate distributions, vol. 1},}\
  \ (\bibinfo  {publisher} {John Wiley \& Sons},\ \bibinfo {address} {New
  York},\ \bibinfo {year} {1994})\ Chap.\ \bibinfo {chapter} {20. Pareto
  Distributions}, pp.\ \bibinfo {pages} {573--627}\BibitemShut {NoStop}%
\bibitem [{\citenamefont {Boettle}\ \emph {et~al.}(2013)\citenamefont
  {Boettle}, \citenamefont {Rybski},\ and\ \citenamefont
  {Kropp}}]{BoettleRK2013}%
  \BibitemOpen
  \bibfield  {author} {\bibinfo {author} {\bibfnamefont {M.}~\bibnamefont
  {Boettle}}, \bibinfo {author} {\bibfnamefont {D.}~\bibnamefont {Rybski}}, \
  and\ \bibinfo {author} {\bibfnamefont {J.~P.}\ \bibnamefont {Kropp}},\
  }\href@noop {} {\bibfield  {journal} {\bibinfo  {journal} {Water Resour.
  Res.}\ }\textbf {\bibinfo {volume} {49}},\ \bibinfo {pages} {1199} (\bibinfo
  {year} {2013})}\BibitemShut {NoStop}%
\bibitem [{\citenamefont {Boettle}\ \emph {et~al.}(2015)\citenamefont
  {Boettle}, \citenamefont {Rybski},\ and\ \citenamefont
  {Kropp}}]{BoettleRK2015}%
  \BibitemOpen
  \bibfield  {author} {\bibinfo {author} {\bibfnamefont {M.}~\bibnamefont
  {Boettle}}, \bibinfo {author} {\bibfnamefont {D.}~\bibnamefont {Rybski}}, \
  and\ \bibinfo {author} {\bibfnamefont {J.~P.}\ \bibnamefont {Kropp}},\
  }\href@noop {} {\bibfield  {journal} {\bibinfo  {journal} {submitted}\ }
  (\bibinfo {year} {2015})}\BibitemShut {NoStop}%
\bibitem [{\citenamefont {Boettle}\ \emph {et~al.}(2016)\citenamefont
  {Boettle}, \citenamefont {Costa}, \citenamefont {Vousdoukas}, \citenamefont
  {Costa}, \citenamefont {Floater}, \citenamefont {Kriewald}, \citenamefont
  {Zhou}, \citenamefont {Chen}, \citenamefont {Kropp},\ and\ \citenamefont
  {Rybski}}]{BoettleRK2016}%
  \BibitemOpen
  \bibfield  {author} {\bibinfo {author} {\bibfnamefont {M.}~\bibnamefont
  {Boettle}}, \bibinfo {author} {\bibfnamefont {L.}~\bibnamefont {Costa}},
  \bibinfo {author} {\bibfnamefont {M.}~\bibnamefont {Vousdoukas}}, \bibinfo
  {author} {\bibfnamefont {H.}~\bibnamefont {Costa}}, \bibinfo {author}
  {\bibfnamefont {G.}~\bibnamefont {Floater}}, \bibinfo {author} {\bibfnamefont
  {S.}~\bibnamefont {Kriewald}}, \bibinfo {author} {\bibfnamefont
  {B.}~\bibnamefont {Zhou}}, \bibinfo {author} {\bibfnamefont {P.-Y.}\
  \bibnamefont {Chen}}, \bibinfo {author} {\bibfnamefont {J.~P.}\ \bibnamefont
  {Kropp}}, \ and\ \bibinfo {author} {\bibfnamefont {D.}~\bibnamefont
  {Rybski}},\ }\href@noop {} {\bibfield  {journal} {\bibinfo  {journal} {in
  prep.}\ } (\bibinfo {year} {2016})}\BibitemShut {NoStop}%
\bibitem [{\citenamefont {Schmidt-Thom{\'e}}\ and\ \citenamefont
  {Klein}(2013)}]{SchmidtThomeK2013}%
  \BibitemOpen
  \bibfield  {author} {\bibinfo {author} {\bibfnamefont {P.}~\bibnamefont
  {Schmidt-Thom{\'e}}}\ and\ \bibinfo {author} {\bibfnamefont {J.}~\bibnamefont
  {Klein}},\ }\href@noop {} {\emph {\bibinfo {title} {Climate Change Adaptation
  in Practice -- From Strategy Development to Implementation}}}\ (\bibinfo
  {publisher} {Wiley-Blackwell},\ \bibinfo {year} {2013})\BibitemShut {NoStop}%
\bibitem [{\citenamefont {Prahl}\ \emph {et~al.}(2012)\citenamefont {Prahl},
  \citenamefont {Rybski}, \citenamefont {Kropp}, \citenamefont {Burghoff},\
  and\ \citenamefont {Held}}]{PrahlRKBH2012}%
  \BibitemOpen
  \bibfield  {author} {\bibinfo {author} {\bibfnamefont {B.~F.}\ \bibnamefont
  {Prahl}}, \bibinfo {author} {\bibfnamefont {D.}~\bibnamefont {Rybski}},
  \bibinfo {author} {\bibfnamefont {J.~P.}\ \bibnamefont {Kropp}}, \bibinfo
  {author} {\bibfnamefont {O.}~\bibnamefont {Burghoff}}, \ and\ \bibinfo
  {author} {\bibfnamefont {H.}~\bibnamefont {Held}},\ }\href@noop {} {\bibfield
   {journal} {\bibinfo  {journal} {Geophys. Res. Lett.}\ }\textbf {\bibinfo
  {volume} {39}},\ \bibinfo {pages} {L06806} (\bibinfo {year}
  {2012})}\BibitemShut {NoStop}%
\bibitem [{\citenamefont {Prahl}\ \emph {et~al.}(2015)\citenamefont {Prahl},
  \citenamefont {Rybski}, \citenamefont {Burghoff},\ and\ \citenamefont
  {Kropp}}]{PrahlRBK2015}%
  \BibitemOpen
  \bibfield  {author} {\bibinfo {author} {\bibfnamefont {B.~F.}\ \bibnamefont
  {Prahl}}, \bibinfo {author} {\bibfnamefont {D.}~\bibnamefont {Rybski}},
  \bibinfo {author} {\bibfnamefont {O.}~\bibnamefont {Burghoff}}, \ and\
  \bibinfo {author} {\bibfnamefont {J.~P.}\ \bibnamefont {Kropp}},\ }\href
  {\doibase 10.5194/nhess-15-769-2015} {\bibfield  {journal} {\bibinfo
  {journal} {Nat. Hazards Earth Syst. Sci.}\ }\textbf {\bibinfo {volume}
  {15}},\ \bibinfo {pages} {769} (\bibinfo {year} {2015})}\BibitemShut
  {NoStop}%
\end{thebibliography}

%

\end{document}